\documentclass[twocolumn,trackchanges]{aastex7}

\shorttitle{AGN jet-induced gas motion in Cygnus A}
\shortauthors{A. Majumder et al.}

\usepackage{hyperref}
\usepackage{placeins}
\usepackage{graphicx}
\usepackage{amsmath}

\begin{document}

\title{Spectrally Resolved Gas Kinematics in Cygnus A: XRISM Detects AGN Jet-induced Velocity Dispersion in Multi-temperature Gas}

\author[orcid=0000-0002-3525-7186]{Anwesh Majumder}
\affiliation{Waterloo Centre for Astrophysics, Department of Physics and Astronomy, 200 University Avenue West, Waterloo,  Ontario N2L 3G1, Canada}
\affiliation{Space Research Organisation Netherlands, Niels Bohrweg 4, Leiden, South Holland 2333 CA, The Netherlands}
\email[show]{anwesh.majumder@uwaterloo.ca}  

\author{T. Heckman}
\affiliation{Center for Astrophysical Sciences, William H. Miller III Department of Physics and Astronomy, Johns Hopkins University, Baltimore, Maryland 21218, USA}
\affiliation{School of Earth and Space Exploration, Arizona State University, Tempe, Arizona 85287, USA}
\email{thechma1@jhu.edu}

\author{J. Meunier} 
\affiliation{Waterloo Centre for Astrophysics, Department of Physics and Astronomy, 200 University Avenue West, Waterloo,  Ontario N2L 3G1, Canada}
\email{jjjmeunier@uwaterloo.ca}

\author[orcid=0000-0002-9714-3862]{A. Simionescu}
\affiliation{Space Research Organisation Netherlands, Niels Bohrweg 4, Leiden, South Holland 2333 CA, The Netherlands}
\affiliation{Leiden Observatory, Leiden University, Niels Bohrweg 2, Leiden, South Holland 2333 CA, The Netherlands}
\affiliation{Kavli Institute for the Physics and Mathematics of the Universe, The University of Tokyo, Kashiwa, Chiba 277-8583, Japan}
\email{a.simionescu@sron.nl}

\author{B.R. McNamara}
\affiliation{Waterloo Centre for Astrophysics, Department of Physics and Astronomy, 200 University Avenue West, Waterloo,  Ontario N2L 3G1, Canada}
\email{mcnamra@uwaterloo.ca}

\author[orcid=0000-0001-9911-7038]{L. Gu}
\affiliation{Space Research Organisation Netherlands, Niels Bohrweg 4, Leiden, South Holland 2333 CA, The Netherlands}
\affiliation{Leiden Observatory, Leiden University, Niels Bohrweg 2, Leiden, South Holland 2333 CA, The Netherlands}
\affiliation{RIKEN High Energy Astrophysics Laboratory, 2-1 Hirosawa, Wako, Saitama 351-0198, Japan}
\affiliation{Department of Physics, Tokyo University of Science, 1-3 Kagurazaka, Shinjuku-ku, Tokyo 162-8601, Japan}
\email{l.gu@sron.nl}

\author[orcid=0000-0001-5655-1440]{A. Ptak}
\affiliation{NASA / Goddard Space Flight Center, 8800 Greenbelt Rd, Greenbelt, MD 20771, USA}
\affiliation{Center for Astrophysical Sciences, William H. Miller III Department of Physics and Astronomy, Johns Hopkins University, Baltimore, Maryland 21218, USA}
\email{ptak@pha.jhu.edu}

\author[orcid=0000-0002-2397-206X]{E. Hodges-Kluck}
\affiliation{NASA / Goddard Space Flight Center, 8800 Greenbelt Rd, Greenbelt, MD 20771, USA}
\email{edmund.hodges-kluck@nasa.gov}

\author[orcid=0000-0001-6366-3459]{M. Yukita}
\affiliation{NASA / Goddard Space Flight Center, 8800 Greenbelt Rd, Greenbelt, MD 20771, USA}
\affiliation{Center for Astrophysical Sciences, William H. Miller III Department of Physics and Astronomy, Johns Hopkins University, Baltimore, Maryland 21218, USA}
\email{myukita1@jhu.edu}

\author[orcid=0000-0002-6470-2285]{M.W. Wise}
\affiliation{Space Research Organisation Netherlands, Niels Bohrweg 4, Leiden, South Holland 2333 CA, The Netherlands}
\affiliation{Anton Pannekoek Institute, University of Amsterdam, Science Park 904, Amsterdam, North Holland 1098 XH, The Netherlands}
\email{m.w.wise@sron.nl}

\author[orcid=0000-0002-4430-8846]{N. Roy}
\affiliation{Center for Astrophysical Sciences, William H. Miller III Department of Physics and Astronomy, Johns Hopkins University, Baltimore, Maryland 21218, USA}
\email{namratar@asu.edu}
 
\begin{abstract}

We report spectral analysis on a 170 ks XRISM \textit{Resolve} exposure of the core of Cygnus A. Analyzing the full field of view spectrum in the $1.7-12.0$ keV band, we find evidence for two-temperature cluster gas. The hotter ($kT = 5.53 \pm 0.13$ keV) gas has a velocity dispersion of $261 \pm 13$ km s$^{-1}$ and a bulk velocity of $120 \pm 20$ km s$^{-1}$ with respect to the central galaxy. The cooler gas ($kT = 2.0^{+0.4}_{-0.3}$ keV) has an even broader velocity dispersion of $440 \pm 130$ km s$^{-1}$, with a systematic uncertainty of $120$ km s$^{-1}$. The relative line-of-sight velocity between the hotter and cooler gas can be as high as $450 \pm 140$ km s$^{-1}$. We interpret the high velocity dispersions as a combination of turbulence and bulk motion due to the cocoon shock. The upper limit on the non-thermal pressure fraction for the hotter gas is $7.7 \pm 0.7$\%. We associate the cooler gas with the central region ($<35$ kpc) and the hotter phase with the gas surrounding it ($35-100$ kpc). The total energy due to the kinetic motion is $5.1 \times 10^{60}$ erg, consistent with the energy associated with the central radio source. The kinetic energy injection rate is $6.9 \times 10^{44}-7.4 \times 10^{45}$ erg s$^{-1}$ under varying assumptions of injection timescales. The range of injection power is higher than the cooling luminosity, and thus the heating and cooling rates in Cygnus A are unbalanced.

\end{abstract}


\section{Introduction} \label{sec:intro}

Feedback from active galactic nuclei (AGN) is a unifying theme in modern astrophysics that is crucial to our understanding of how the AGN and the baryonic matter surrounding it evolves over cosmic time \citep{ferrarese_fundamental_2000,fabian_observational_2012,mcnamara_mechanical_2012,eckert_feedback_2021}. Jet-mode feedback, in particular, is the most critical mechanism that shapes this evolution due to the enormous energy injected over a large spatial extent  \citep{kondapally_cosmic_2023,heckman_global_2023,heckman_mergers_2024}. Jets can inject energy of $10^{58}-10^{62}$ ergs into the hydrostatic atmosphere that results in shocks (e.g., \citealt{fabian_deep_2003,simionescu_large-scale_2009,randall_very_2015,snios_cocoon_2018}), cavities (e.g., \citealt{birzan_systematic_2004,rafferty_feedback-regulated_2006,panagoulia_volume-limited_2014}), sound waves (e.g., \citealt{fabian_deep_2003,forman_filaments_2007}) and turbulence (e.g., \citealt{zhuravleva_turbulent_2014,2016Natur.535..117H,2018PASJ...70...10H,rose_xrism_2025}). One of the key open questions in cluster physics is what the energy partition is between these different dissipation mechanisms. 

Answering how much turbulence contributes to the total energy budget remained uncertain due to the difficulty of directly measuring jet-driven gas motions of few hundred km s$^{-1}$ \citep{gaspari_mechanical_2012,valentini_agn-stimulated_2015,gaspari_shaken_2018} from low spectral resolution \textit{Chandra} (ACIS) and \textit{XMM-Newton} (EPIC) data ($\Delta E \sim 150$ eV at 6.4 keV).\footnote{\href{https://xmm-tools.cosmos.esa.int/external/xmm_user_support/documentation/uhb/basics.html}{https://xmm-tools.cosmos.esa.int/external/xmm\_user\_support/\\documentation/uhb/basics.html}}$^,$\footnote{\href{https://cxc.harvard.edu/cal/Acis/}{https://cxc.harvard.edu/cal/Acis/}} Instruments like the reflection grating spectrometer (RGS) onboard \textit{XMM-Newton} were more succcessful in this endeavor ($\Delta E \sim 2-3$ eV at 1 keV)$^{11}$ and upper limits as high as 500 km s$^{-1}$ were placed in many clusters (\citealt{pinto_chemical_2015}; also see \citealt{pinto_lessons_2019} for a review). However, slitless spectrometers like RGS change properties of spectral lines (position and width) arising from regions offset from the dispersion axis. For extended objects like galaxy clusters, this effect can thus introduce systematic uncertainty on derived turbulent velocity. The recently launched XRISM telescope \citep{tashiro_x-ray_2025}, the successor of \textit{Hitomi} \citep{2014SPIE.9144E..25T}, gets around this problem by introducing non-dispersive high-resolution spectroscopy using the \textit{Resolve} instrument onboard ($\Delta E \sim 5$ eV at 6.4 keV; \citealt{2022SPIE12181E..1SI}). 

Beyond non-dispersive high spectral resolution requirements, constraining the AGN-driven jets' role in driving gas motion is further complicated by the fact that very different assumptions regarding energy injection can lead to similar predictions. For example, varying assumptions used in previous studies by \cite{hillel_hitomi_2017}, \cite{lau_physical_2017}, \cite{li_agn_2017} were able to explain the \textit{Hitomi} gas motion measurements of the Perseus cluster \citep{2016Natur.535..117H}. It can also be difficult to isolate AGN jet-driven motions from sloshing or minor mergers \citep{ascasibar_origin_2006,ichinohe_substructures_2019}. It is also unclear whether turbulence from AGN jets alone can inject enough energy to stop the ICM gas from cooling. In some clusters like Perseus and Virgo, it has been suggested that the turbulent heating rate is just enough to match the radiative cooling rate \citep{zhuravleva_turbulent_2014} while in other clusters it has been suggested that the turbulence from AGN-driven jets will struggle to replenish the lost energy \citep{bambic_limits_2018,2025Natur.638..365X,rose_xrism_2025,fujita_xrism_2025}. In light of this, it is critical to measure gas velocities spanning a range of spatial scales,  AGN power, and dynamical states to gain a deeper understanding on jet-driven gas motion. 

\begin{figure*}
    \centering
    \includegraphics[width=\textwidth, keepaspectratio]{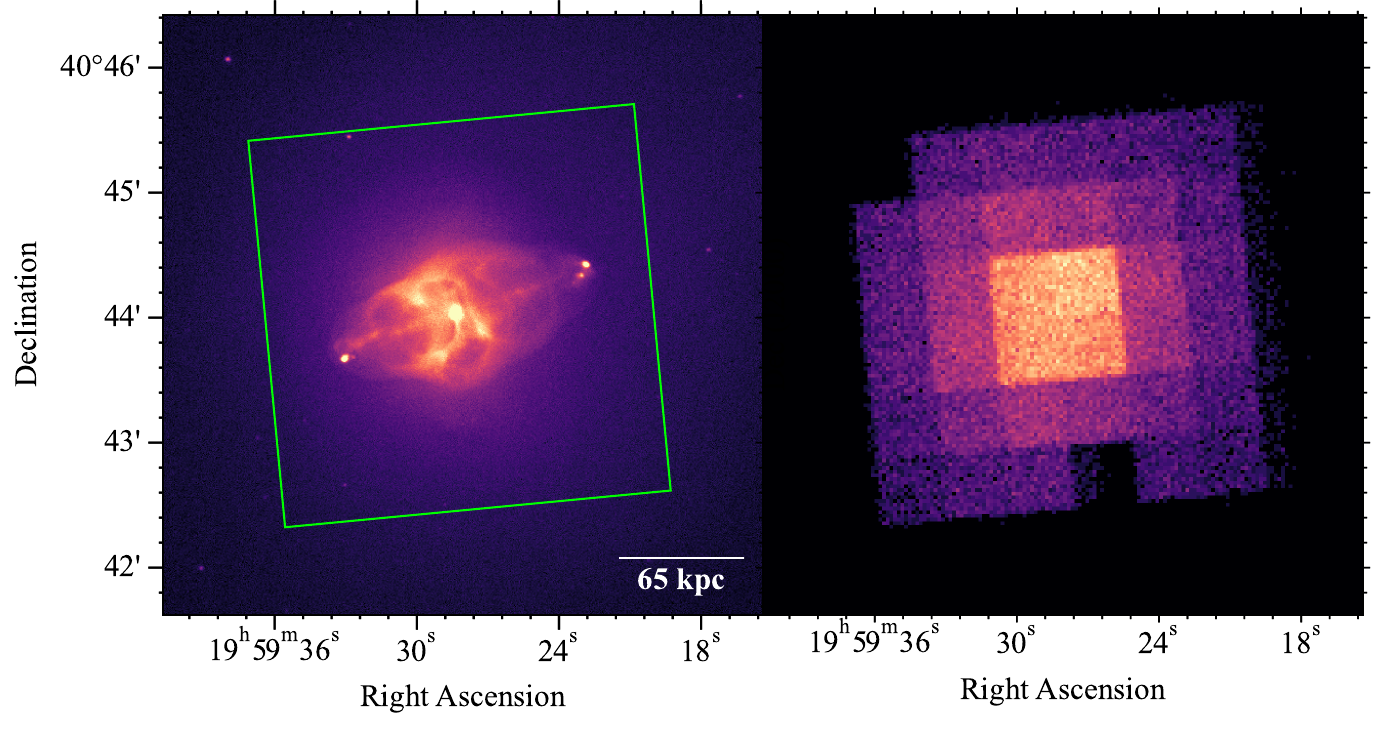}
    \caption{\small{
\emph{Left:} 2.2 Ms \textit{Chandra} image of the core of Cygnus A in the $0.7-7.0$ keV band. The image was processed according to the description in \cite{majumder_decoding_2024}. The \textit{Resolve} field of view (FOV) has been overlaid in green. \emph{Right:} \textit{Resolve} view of the same region with pixels 12 and 27 removed. The image was created in the $1.7-12.0$ keV band. Only the Hp events were included.}}
\label{fig:fov}
\end{figure*}

XRISM has recently made progress in exploring this parameter space by providing insight into the AGN feedback process for the Centaurus cluster \citep{2025Natur.638..365X}, A2029 \citep{xrism_collaboration_xrism_2025-1}, Ophiuchus cluster \citep{fujita_xrism_2025} and Hydra A \citep{rose_xrism_2025}. Centaurus has a low jet power (P $\sim 7 \times 10^{42}$ erg s$^{-1}$; \citealt{rafferty_feedback-regulated_2006}),  while Ophiuchus currently have little active energy input from the AGN \citep{fujita_xrism_2025}. A2029 has a jet power of $3 \times 10^{44}$ erg s$^{-1}$ \citep{xrism_collaboration_xrism_2025-1} while Hydra A have a similar power inputs in the core ($2 \times 10^{44}$ erg s$^{-1}$; \citealt{wise_xray_2007}). In all of these cases, the measured line-of-sight velocity dispersion is less than 170 km s$^{-1}$. These results so far suggest that the effect of jet-driven gas motion is limited below a power of about a few times $10^{44}$ erg s$^{-1}$. However, it is unknown whether this result holds as the jet power increases. 

In this work, we undertake an investigation of gas motion for a cluster with a high jet power ($\geq 10^{46}$ erg s$^{-1}$). We analyze a 170 ks XRISM observation of the well-known radio galaxy Cygnus A. Cygnus A is a Fanaroff-Riley class II radio galaxy \citep{1974MNRAS.167P..31F} that is both nearby ($z \sim 0.056$; \citealt{1977ApJ...217...45S,owen_cluster_1997}) and $10^3$ times brighter than any other radio sources at a similar distance \citep{1996cyga.book.....C}. The cluster embedding Cygnus A is a dense ($n_e \sim 10^{-2}$ cm$^{-3}$), hot ($kT \sim 5-7$ keV) environment \citep{markevitch_physics_1999,sarazin_merger_2013,snios_cocoon_2018,majumder_decoding_2024} which is interacting with the radio lobes from Cygnus A and is energetically coupled to it \citep{1984MNRAS.211..981A,1994MNRAS.270..173C}. The jet is driving a cocoon shock ($M \sim 1.2 - 1.7$) and is carrying up to $10^{46}$ erg s$^{-1}$ of power into the cluster gas \citep{wilson_cavity_2006,snios_cocoon_2018}. Moreover, prominent cavities have been clearly seen due to the interaction between the radio lobes and the intracluster medium (ICM) \citep{1994MNRAS.270..173C,smith_chandra_2002,wilson_cavity_2006,chon_discovery_2012}. Such an extreme environment, along with shock velocities reported to be as high as 2000 km s$^{-1}$ \citep{snios_cocoon_2018} can justifiably lead to larger velocity dispersion compared to previous results in the literature. In the following sections, we demonstrate that this is indeed the case.

We begin in \S\ref{sec:obs_data} with a description of the observation and the data reduction process. We then describe in detail the spectral extraction method in \S\ref{sec:spec_ex}, followed by a discussion of the observed full FOV spectrum in \S\ref{sec:spectrum}. The spectrum contains emission from both the cluster and the central AGN, and thus the spectral analysis setup is discussed thoroughly in \S\ref{sec:spec_fit}. We then describe our result for a single temperature and two temperature case in \S\ref{sec:results}. We conclude by discussing the implications of our work in \S\ref{sec:discussion}, followed by a summary in \S\ref{sec:summary}.

We have assumed a $\Lambda$CDM cosmology with $H_0 = 70$ km s$^{-1}$ Mpc$^{-1}$, $\Omega_m = 0.3$ and $\Omega_{\Lambda} = 0.7$.  The corresponding linear scale at the redshift of Cygnus A is 65 kpc per arcmin. All errors reported in this work are of $1\sigma$ significance.

\section{Observation and Data Reduction} \label{sec:obs_data}

The core of Cygnus A was observed with XRISM from 6th October 2024 to 10th October 2024 (ObsID: 201120010) with an aimpoint of RA = $19^{\textrm{\scriptsize{h}}} 59^{\textrm{\scriptsize{m}}} 29^{\textrm{\scriptsize{s}}}.36$ and Dec = $40^{\circ} 44' 01''.6$ for a total net exposure of 170 ks. For this study, we only use data from \textit{Resolve} $-$ a 3.1' $\times$ 3.1' microcalorimeter array containing $6 \times 6$ pixels. This covers an area of roughly 200 x 200 kpc in Cygnus A. Out of these pixels, one pixel (pixel 12) is used for calibration. The rest of the pixels each produce spectra from incident X-rays with an energy resolution of 4.5 eV \citep{2024SPIE13093E..1KP}. In our observation, XRISM's Beryllium window gate valve was closed and therefore, soft energy photons ($<1.7$ keV) are heavily absorbed. As such, \textit{Resolve} is only sensitive above 1.7 keV. 

We reprocessed the data from the \textit{Resolve} detector using the XRISM team's Build 8 software and applied calibration from CALDB v8. The event file was filtered according to \texttt{RISE\_TIME} parameter by applying the criterion \texttt{((((RISE\_TIME+0.00075*DERIV\_MAX)>46) \
\&\&((RISE\_TIME+0.00075*DERIV\_MAX)<58))\&\&ITYPE<4)} \texttt{||(ITYPE==4))\&\&STATUS[4]==b0}. The standard `\texttt{RISE\_TIME}' criterion allows us to screen for non-X-ray events while \texttt{STATUS[4]} allows us to screen for false-positive frame events. All pixels were included apart from the calibration pixel 12 and additionally, pixel 27. Pixel 27 has shown unexpected gain jumps not captured by the standard calibration cadence. We used only high primary (Hp) events (highest energy resolution) for spectral analysis by setting \texttt{ITYPE==0}. Almost $97\%$ of all events were Hp events in the $1.7-12.0$ keV band (image shown in Figure \ref{fig:fov}).

The energy scale error for the calibration pixel was $\sim$ 0.022 eV.\footnote{\href{https://heasarc.gsfc.nasa.gov/FTP/xrism/postlaunch/gainreports/2/201120010_resolve_energy_scale_report.pdf}{201120010\_resolve\_energy\_scale\_report.pdf}} Adding the current energy scale accuracy of $< 0.3$ eV \footnote{\href{https://tinyurl.com/5chvzhu5}{Resolve instrument}} in quadrature results in a total bulk velocity systematic uncertainty of $\sim 15$ km s$^{-1}$ at 6 keV. The current line-spread function accuracy of 0.17 eV between $6-7$ keV, on the other hand, results in a systematic uncertainty of $< 1$ km s$^{-1}$ for $> 260$ km s$^{-1}$ velocity dispersion.

XRISM had a barycentric velocity of $-13$ km s$^{-1}$ with respect to Cygnus A at the epoch of observation (JD 2460590). This velocity was derived using the method described in \cite{wright_barycentric_2014}\footnote{\href{https://astroutils.astronomy.osu.edu/exofast/barycorr.html\#}{https://astroutils.astronomy.osu.edu/exofast/barycorr.html}}, as well as accounting for $\sim$8 km s$^{-1}$ low-Earth orbit velocity. The barycentric motion has been corrected for in all reported redshifts. 

\section{Spectral Extraction} \label{sec:spec_ex}

We extracted spectra from the full array of \textit{Resolve} apart from pixels 12 and 27. We then used the latest calibration file `\texttt{xa\_rsl\_rmfparam\_20190101v006.fits}' to create the redistribution matrix file (RMF). We used the X-sized RMF for all our analysis. The X-size RMF matrix models the secondary response components most accurately, albeit at the expense of convolution speed. 


For the ancillary response file (ARF), we used a point-source ARF to model emission from the central AGN. The point source was centered at the aimpoint for the ARF generating task \texttt{xaarfgen}. For the extended cluster, we used a 5 arcmin $\times$ 5 arcmin \textit{Chandra} image centered on Cygnus A (RA = $19^{\textrm{\scriptsize{h}}} 59^{\textrm{\scriptsize{m}}} 28^{\textrm{\scriptsize{s}}}.36$, Dec =  $40^{\circ} 44' 02''.1$; \citealt{fey_second_2004}) in the $0.7-7.0$ keV band as the source model input for \texttt{xaarfgen}. This was done to properly take into account scattering of photons in and out of the field of view due to XRISM's large point-spread function. The \textit{Chandra} image of Cygnus A was created according to the procedure described in \cite{majumder_decoding_2024}. While the extended ARF changes normalization of the spectral components, it does not affect parameters like velocity broadening or redshift measurements. 

The non-X-ray background (NXB) was extracted from a database of \textit{Resolve} night-Earth data using the task \texttt{rslnxbgen}. We used standard filters as recommended by the XRISM team.\footnote{\href{https://heasarc.gsfc.nasa.gov/docs/xrism/analysis/nxb/resolve_nxb_db.html}{https://heasarc.gsfc.nasa.gov/docs/xrism/analysis/nxb/\\resolve\_nxb\_db.html}} The NXB was modeled independently and before analyzing the dataset, specifically within the 1.7–17 keV energy range. This model includes a power-law component along with Gaussian profiles for detector emission lines. Initially, the overall NXB normalization was fitted for each spectrum, after which the individual emission line normalizations were adjusted. The finalized NXB model was then fixed and used in all following spectral fits.

\section{The Spectrum} \label{sec:spectrum}

As can be seen from Figure \ref{fig:1t_fit}, we present the full FOV spectra of Cygnus A along with the NXB model. From the bottom right subfigure, the Fe XXV $K_{\alpha}$ ($\sim$6.3 keV redshifted), XXVI ($\sim$6.55 keV redshifted) and XXV $K_{\beta}$, along with Ni lines($\sim$7.25-7.45 keV redshifted) from the hot cluster are clearly visible. The presence of the rest-frame 6.4 keV ($\sim$6.06 keV redshifted) Fe fluorescence line from the central AGN \citep{young_chandra_2002} is also obvious.  This line feature is due to the irradiation of low- and intermediate-ionized Fe atoms in the circumnuclear gas by the hard X-ray continuum of the AGN. The data at the higher energies ($>8$ keV) in the $1.7-12.0$ keV plot also resemble a power-law continuum from the central AGN rather than a bremsstrahlung cutoff expected from cluster emission.

It is clear that detailed modeling of emission from both the central AGN and the ICM is necessary to fully interpret the richness of the data. In this paper, we explore the ICM, and in an upcoming paper (Majumder et al., in prep.), we interpret the AGN spectrum. In the following section, we describe all models used to interpret the data.

\section{Spectral Fitting and Modeling} \label{sec:spec_fit}

\subsection{Preparing the Spectrum for Fitting}

We used \texttt{SPEX v3.08.01} for all our spectral fitting \citep{1996uxsa.conf..411K,2018zndo...2419563K,2020zndo...4384188K}.\footnote{\href{https://spex-xray.github.io/spex-help/index.html}{https://spex-xray.github.io/spex-help/index.html}} The full FOV spectrum, RMF, the point-source and the extended source ARF were converted to the \texttt{SPEX} format \citep{kaastra_optimal_2016} using the \texttt{SPEX} task \texttt{trafo}. The response file in the new format was then optimally binned according to energy resolution \citep{kaastra_optimal_2016} with the \texttt{SPEX} command \texttt{rbin} in the $1.7-12.0$ keV band. After optimally binning, we again used \texttt{trafo} to load the spectrum, RMF, the extended source ARF in sector one in SPEX format and the same spectrum, RMF, but point-source ARF in sector two (see \citealt{kaastra_optimal_2016} for a discussion on sectors).\footnote{\href{https://spex-xray.github.io/spex-help/theory/fitting/sectors.html}{https://spex-xray.github.io/spex-help/theory/fitting/sectors.html}} Arranging the data in this way ensures that we can fold models for the cluster and the AGN through different ARFs. 

For all our fits, we used the C-statistic for minimization \citep{1979ApJ...228..939C,kaastra_use_2017}. All abundances were measured with respect to the reference proto-solar abundance table of \cite{2009LanB...4B..712L}. 

\subsection{Spectral Modeling} \label{subsec:spec_model}

The Cygnus A full FOV spectrum consists of features from both ICM and the central AGN. Therefore, we constructed spectral models for each source. We describe the ICM model below and detail the free parameters in Table \ref{tab:ICM_fits}. The description of AGN models, along with their fitted values, will be detailed in an upcoming work (Majumder et al., in prep.). We therefore only give a brief overview of the AGN model below. In all our fits, the distance to Cygnus A was set to a redshift of $z \approx 0.0561$ \citep{owen_cluster_1997}. We do not take into account the cosmic X-ray background since the emission from the cluster containing Cygnus A is more than two orders of magnitude brighter than the X-ray background within the \textit{Resolve} field of view (see Figure 3 of \citealt{majumder_decoding_2024}). 

\subsubsection{The ICM Model}

We modeled the ICM first with one collisional ionization equilibrium (\texttt{cie}) component, followed by two \texttt{cie} models corresponding to hotter and cooler gas in the cluster. In both cases, the normalization, temperature, and velocity broadening of any \texttt{cie} component were allowed to vary independently. For the two \texttt{cie} case, the Si, S, Ar, Ca and Fe abundances of the hotter gas were also allowed to vary, but abundances of the corresponding elements of the cooler gas were coupled to those of the hotter gas. All remaining elements from C to Ni were coupled to the Fe abundance. The ion temperature and electron temperature of each \texttt{cie} component were also coupled to each other.

We redshifted each \texttt{cie} component using the \texttt{reds} model of \texttt{SPEX}. Independently varying the redshift for each component in the two \texttt{cie} case allows us to investigate any systematic velocity shift between the hotter and cooler gas. Any velocity shift of either \texttt{cie} component with respect to the central galaxy can also be investigated in this way. 

Finally, the emission from the ICM was absorbed through the Milky Way galactic absorption column of $n_H = 4.14 \times 10^{21}$ atoms cm$^{-2}$. The weighted hydrogen absorption column was calculated using the method proposed by \cite{willingale_calibration_2013}.\footnote{\href{https://www.swift.ac.uk/analysis/nhtot/}{https://www.swift.ac.uk/analysis/nhtot/}} We used the \texttt{hot} model in \texttt{SPEX} for absorbing the \texttt{cie} components. The temperature of the \texttt{hot} model was fixed to $10^{-6}$ keV for neutral absorption. In order to take into account any systematics with ARF calibration in the soft band, we allowed the $n_H$ to vary in our fits.

\subsubsection{The AGN Model}

We modeled the non-thermal emission from the central AGN with a power-law using the \texttt{pow} model in \texttt{SPEX}. The normalization and the spectral index were allowed to vary freely. The 6.4 keV Fe line was modeled using laboratory measurements of the line \citep{1997PhRvA..56.4554H}. This line model was then folded through multiple \texttt{spei} models of SPEX to simulate reflection by circumnuclear material at various distances from the central black hole. The distance where this reflection takes place was allowed to vary. The inclination angles of the \texttt{spei} models were coupled together and allowed to vary. The emissivity index was set to $q=3$ for an isotropic source and a flat disk. A non-spinning black hole was also assumed for the fits. In addition, to investigate any ionized outflowing winds, we used the \texttt{pion} model to simulate an ionized absorber with a bulk velocity. We assumed a covering fraction $f_{\textrm{\scriptsize{cov}}} = 1$. We then varied the average systematic velocity of the absorber, $z_v$, the root mean square velocity of ions in the absorber, $v$, the ionized absorption column $N_{\textrm{\scriptsize{ionized}}}$ and the ionization parameter $\xi$.

All AGN components were then absorbed with a thick neutral hydrogen absorption column using the \texttt{hot} model in SPEX. The temperature of the \texttt{hot} model was again fixed to $10^{-6}$ keV for neutral absorption. Finally, the AGN components were shifted with a redshift ($z_{\textrm{\scriptsize{AGN}}}$) using the \texttt{reds} component. This redshift was allowed to vary to independently determine the redshift of the central AGN from the X-ray spectrum. We provide more details in an upcoming work (Majumder et al., in prep.).

\subsection{Ignored Spectral Lines During Fitting}

Our fits ignored the strong Fe XXV-$w$ line as this line can be suppressed due to resonance scattering \citep{1987SvAL...13....3G} not taken into account by the \texttt{cie} models. We discuss resonance scattering in \S\ref{subsec:resonance_scattering}. We also removed the Si He-$\beta$ line from our fits by removing spectral data points in the $2.220-2.265$ keV band. At Cygnus A's redshift, this line fell at the Si edge of the effective area at this energy and thus could not be fit properly. To avoid biasing velocity widths, we therefore removed the line (see \S\ref{appendix:ig_lines} for more details). Finally, we removed the NXB Si fluorescence line by removing data points between $1.735-1.750$ keV band. Despite following standard NXB reduction and modeling procedures, this line could not be fit well presumably due to lingering calibration issues (see \S\ref{appendix:ig_lines} for more details). With this setup, we explored the spectral properties of the ICM and the AGN.

\begin{figure*}
    \centering
    \includegraphics[width=\textwidth, keepaspectratio]{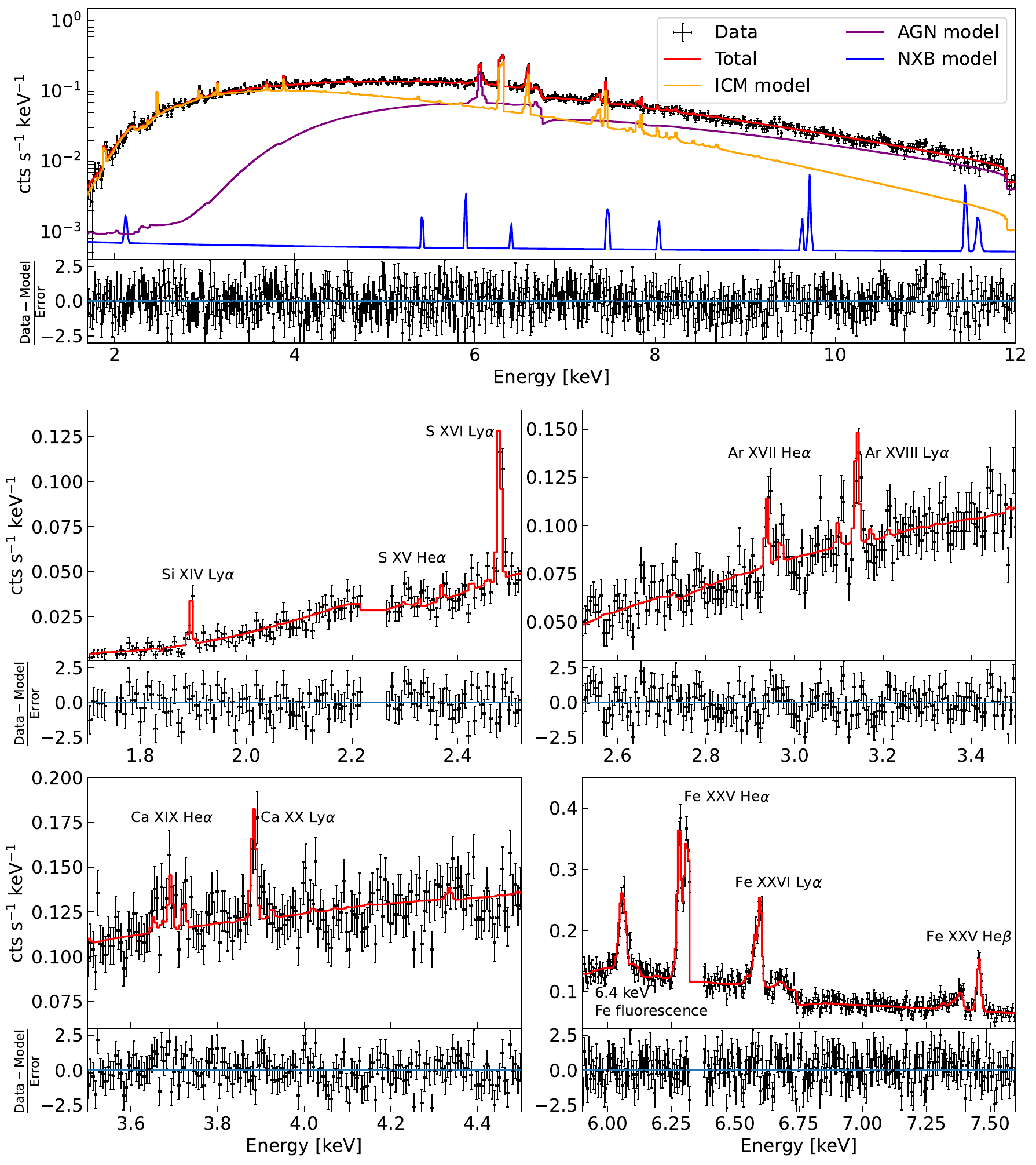}
    \caption{\small{A single \texttt{cie} fit to the Full FOV \textit{Resolve} spectrum of Cygnus A. \emph{Top:} The AGN+ICM model fit to the $1.7-12.0$ keV band data along with the residuals. We also show the NXB model used for the fit. The spectrum has been binned by a factor of 5 in this energy range for visual purposes. \emph{Middle left:} The same spectrum zoomed in the $1.7-2.5$ keV band showing Si and S lines along with the residuals. \emph{Middle right:} The $2.5-3.5$ keV band showing Ar lines along with the residuals. \emph{Bottom left:} The $3.5-4.5$ keV band showing Ca lines along with the residuals. \emph{Bottom right:} The $5.9-7.6$ keV band showing Fe lines from AGN ($6.05$ keV redshfited) and from the ICM. The Middle and bottom panels have been binned by a factor of 2 for visual purposes.}}
\label{fig:1t_fit}
\end{figure*}

\section{Results} \label{sec:results}

\subsection{Single \texttt{cie} Model}

We first fitted one \texttt{cie} model to our data and constrained the ICM parameters. The resulting best-fit is shown in Figure \ref{fig:1t_fit} and the parameters are reported in Table \ref{tab:ICM_fits}. By comparing the obtained \textit{Resolve} redshift with the heliocentric optical redshift of the central galaxy ($z_{\textrm{\scriptsize{CG}}} = 0.05608 \pm 0.00007$; \citealt{owen_cluster_1997}), we conclude that the ICM is redshifted. From these redshifts, we can calculate the line-of-sight bulk velocity as

\begin{equation}   \label{eq:bulk_velocity}
    v_{\textrm{\scriptsize{bulk, ICM-CG}}} = \frac{c (z_{\textrm{\scriptsize{ICM}}} - z_{\textrm{\scriptsize{CG}}})}{1 + z_{\textrm{\scriptsize{CG}}}},
\end{equation}
\noindent where $z_{\textrm{\scriptsize{ICM}}}$ is the ICM redshift and $c$ is the speed of light. The redshifts suggest that the ICM is moving away from the central galaxy with a $v_{\textrm{\scriptsize{bulk, ICM-CG}}} = 100 \pm 20$ km s$^{-1}$ . On the other hand, the line broadening due to gas motions is $\sigma_{\textrm{\scriptsize{1D, 1T}}} = 273 \pm 13$ km s$^{-1}$. This line broadening is one of the largest published so far among the galaxy clusters observed by XRISM \citep{2025Natur.638..365X,xrism_collaboration_xrism_2025-1,xrism_collaboration_xrism_2025,rose_xrism_2025,fujita_xrism_2025}. The line shifts and high broadening are likely due to a combination of turbulence as well as the expanding cocoon observed in Cygnus A \citep{wilson_cavity_2006,snios_cocoon_2018}. 

Despite the C-statistic being within the expected range (Table \ref{tab:ICM_fits}; see \citealt{kaastra_use_2017} for an explanation of expected C-statistic), it can be seen from Figure \ref{fig:1t_fit} and Figure \ref{fig:hot_cool} (left panel) that the wings of certain prominent lines (S XVI Ly$\alpha$ and Ar XVII He$\alpha$) are poorly fitted. Furthermore, the Si and S to Fe ratios from Table \ref{tab:ICM_fits} are strongly super-solar. Since such effects are normally caused by an additional gas component, adding a second \texttt{cie} to our model is pertinent.

\begin{figure*}
    \centering
    \includegraphics[width=\textwidth, keepaspectratio]{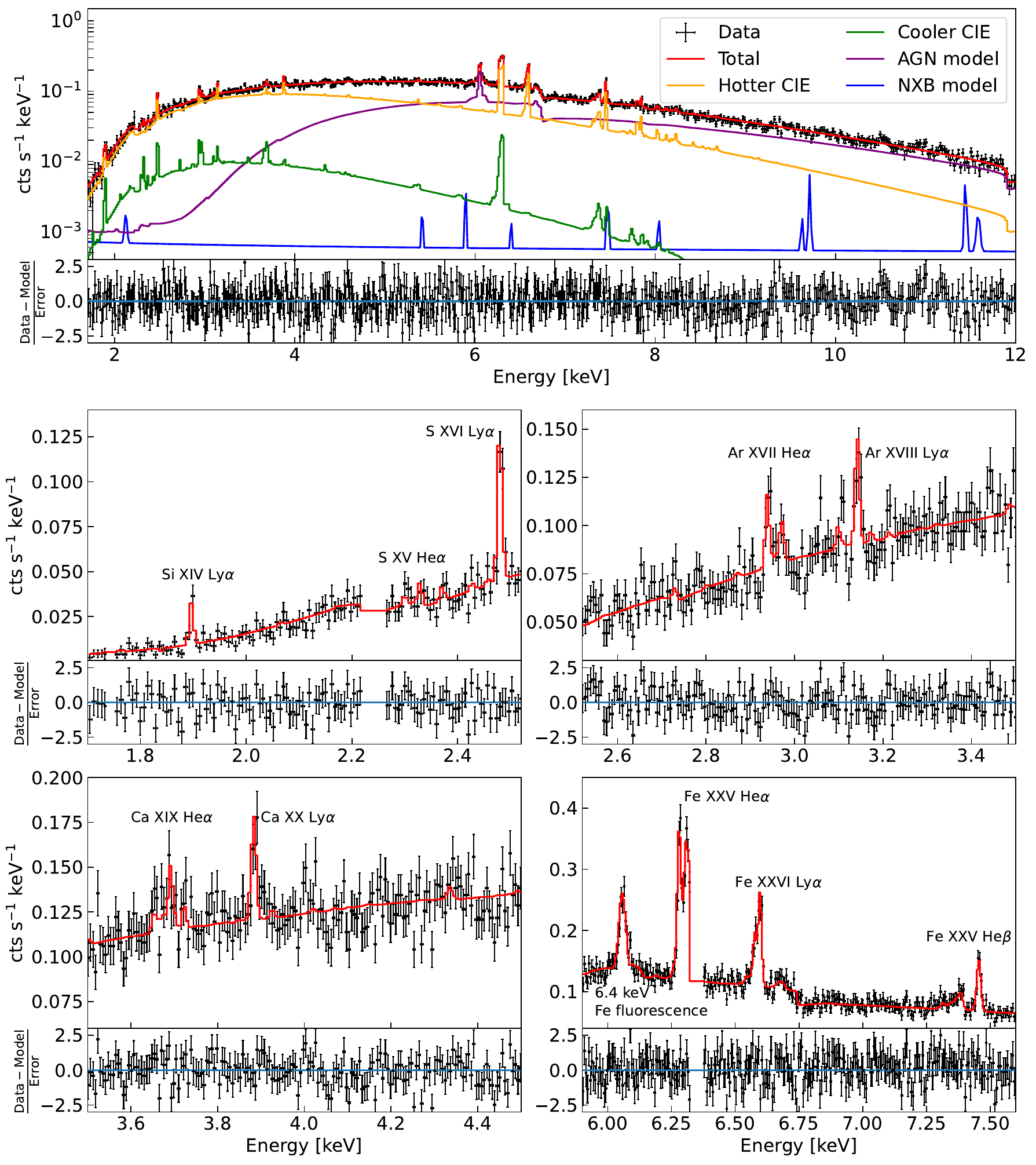}
    \caption{\small{Two \texttt{cie} fit to the Full FOV \textit{Resolve} spectrum of Cygnus A. \emph{Top:} The AGN+ICM model fit to the $1.7-12.0$ keV band data along with the residuals. We also show the NXB model used for the fit. The spectrum has been binned by a factor of 5 in this energy range for visual purposes. \emph{Middle left:} The same spectrum zoomed in the $1.7-2.5$ keV band showing Si and S lines along with the residuals. \emph{Middle right:} The $2.5-3.5$ keV band showing Ar lines along with the residuals. \emph{Bottom left:} The $3.5-4.5$ keV band showing Ca lines along with the residuals. \emph{Bottom right:} The $5.9-7.6$ keV band} showing Fe lines from AGN ($6.05$ keV redshfited) and from the ICM. The Middle and bottom panels have been binned by a factor of 2 for visual purposes.}
\label{fig:2t_fit}
\end{figure*}

\subsection{Two \texttt{cie} Model} \label{subsec:two_cie}

\begin{figure*}
    \centering
    \includegraphics[width=\textwidth, keepaspectratio]{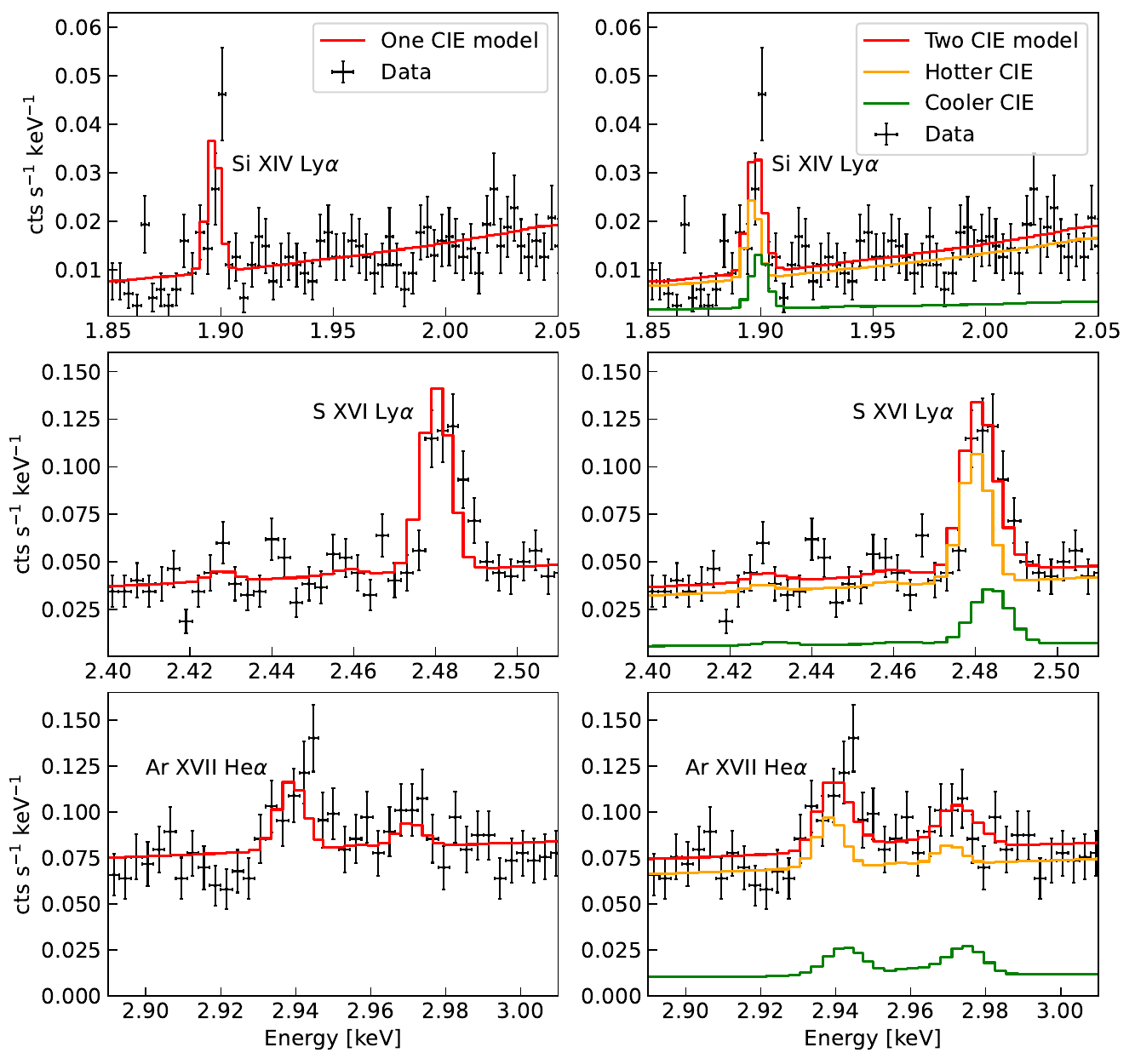}
    \caption{\small{The various lines in the soft band ($1.7-4.5$ keV) causing the broad velocity of cooler component are shown. \emph{Left:} One \texttt{cie} model fit. Wings and strength of these line complexes are fitted poorly. \emph{Right:} Two \texttt{cie} model fit. The yellow line shows the hotter \texttt{cie} component and the green line shows the cooler \texttt{cie} component. The red line shows the total two \texttt{cie} model. The wings and strength of the lines are much better fitted. A shift between hotter and cooler \texttt{cie} component is evident. The cooler component is also much broader than the hotter component.}}
\label{fig:hot_cool}
\end{figure*}

We next added a second \texttt{cie} model to separate the effect of cooler gas from the hotter gas. We show these fits in Figure \ref{fig:2t_fit} and report the fit parameters in Table \ref{tab:ICM_fits}. As we can see from the soft band ($1.7-4.5$ keV; also see right panel of Figure \ref{fig:hot_cool}), the S XVI Ly$\alpha$ and Ar XVII He$\alpha$ lines are better reproduced, and the difference in C-statistic is $\Delta C = 22$. The normalization and temperature of the cooler gas are also constrained with a significance $>3\sigma$. We furthermore calculate the Akaike information criterion (AIC) to understand whether one \texttt{cie} or two \texttt{cie} model is favored. The AIC is defined as

\begin{table}
    \centering
\caption{Best-fit parameters for the core of Cygnus A in the $1.7-12.0$ keV band.} 
\label{tab:ICM_fits}
    \begin{tabular}{cc}
    \hline
    \hline
    Parameter&Value\\
    \tableline
    \multicolumn{2}{c}{\texttt{reds(hot(cie))} + AGN models \tablenotemark{a}}\\
         ICM Redshift &  ($5.643 \pm 0.005) \times 10^{-2}$\\
         Norm ($n_e n_p V$)&$(3.5 \pm 0.1)\times 10^{67}$ cm$^{-3}$\\
        kT&$5.33 \pm 0.11$ keV\\
        $\sigma_{1D}$\tablenotemark{b}&$273 \pm 13$ km s$^{-1}$\\
        Si&$1.4 \pm 0.3$ $Z_{\odot}$\\
        S&$0.88 \pm 0.09$ $Z_{\odot}$\\
        Ar&$0.77 \pm 0.12$ $Z_{\odot}$\\
        Ca &$0.76 \pm 0.10$ $Z_{\odot}$\\
        Fe &$0.57 \pm 0.02$ $Z_{\odot}$\\
 C-stat (Expected C-stat)&$3074$ ($3051 \pm 78$)\\
 \tableline
    \multicolumn{2}{c}{\texttt{reds(hot(cie)) + reds(hot(cie))} + AGN models\tablenotemark{a}}\\
        Hotter gas redshift&  $(5.649 \pm 0.005) \times 10^{-2}$\\
        Hotter gas norm ($n_e n_p V$)&$ (3.1 \pm 0.2)  \times 10^{67}$ cm$^{-3}$\\
        Hotter gas kT&$5.53 \pm 0.13$ keV\\
        Hotter gas $\sigma_{1D}$\tablenotemark{b}&$261 \pm 13$ km s$^{-1}$\\
 Cooler gas redshift&$(5.49 \pm 0.05) \times 10^{-2}$\\
 Cooler gas norm ($n_e n_p V$)&$0.7^{+0.3}_{-0.2} \times 10^{67}$ cm$^{-3}$\\
 Cooler gas kT&$2.0^{+0.4}_{-0.3}$ keV\\
 Cooler gas $\sigma_{1D}$\tablenotemark{b}&$440 \pm 130$ km s$^{-1}$\\
        Si &$1.0^{+0.3}_{-0.2}$ $Z_{\odot}$\\
        S &$0.75 \pm 0.08$ $Z_{\odot}$\\
        Ar &$0.75 \pm 0.11$ $Z_{\odot}$\\
        Ca &$0.78 \pm 0.11$ $Z_{\odot}$\\
        Fe&$0.60 \pm 0.02$ $Z_{\odot}$\\
        C-stat (Expected C-stat)&$3052$ ($3051 \pm 78$)\\
    \tableline
    \end{tabular}
    \tablecomments{$n_e$ is the electron density, $n_p$ is the proton density and $V$ is the line-of-sight enclosed volume. All reported $\sigma_{1D}$ are velocity due to gas motions. The errors reported in this table are statistical.}
   \tablenotetext{a}{The AGN models will be covered in an upcoming work (Majumder et al., in prep.). Only the ICM model parameters are reported in this table.}
   \tablenotetext{b}{In SPEX \texttt{cie} models, $\sigma_{1D}$ is referred to as $v_{\textrm{\scriptsize{RMS}}}$.}
    
\end{table}

\begin{equation} \label{eq:aic}
    \textrm{AIC} = 2k - 2\textrm{ln}(\mathcal{L}),
\end{equation}

\noindent where $k$ is the number of parameters and $\mathcal{L}$ is the likelihood. Since C-statistic is defined as $-2 \times$ Poissonian log-likelihood (see \citealt{1979ApJ...228..939C} and analogous definition in \citealt{kaastra_use_2017}), Equation \ref{eq:aic} becomes

\begin{equation}
    \textrm{AIC} = 2k + \textrm{C-stat}.
\end{equation}

\noindent We can calculate AIC for our data where $k = 23$, $\textrm{C-stat} = 3074$ for one \texttt{cie} + AGN models and $k = 27$, $\textrm{C-stat} = 3052$ for two \texttt{cie} + AGN models (Majumder et al., in prep.). For the one \texttt{cie} model we get $\textrm{AIC} = 3120$ and for two \texttt{cie} models, we get $\textrm{AIC} = 3106$, i.e. $\Delta \textrm{AIC} = 14$. A $\Delta \textrm{AIC}> 10$ is generally considered significant. It is therefore suggestive that the two \texttt{cie} model is a better fit for our data than the one \texttt{cie} model.

Based on our fits, we find that the cooler gas normalization is $\sim$20\% of the hotter gas. The fits also indicate that the hotter and the cooler gas in the ICM have different kinematics. Although the velocity dispersion of the hotter gas remains unchanged within statistical errors ($\sigma_{\textrm{\scriptsize{1D, hot}}} = 261 \pm 13$ km s$^{-1}$), we see that the cooler gas has a dispersion of $\sigma_{\textrm{\scriptsize{1D, cool}}} = 440 \pm 130$ km s$^{-1}$ ($3.4\sigma$ significance). This result suggests that the cooler gas is being affected much more than the hotter gas by the jet. The fit results also indicate that the redshift differences between the hotter and cooler gas are significant. These results can be visually seen in Figure \ref{fig:hot_cool}, where we show the lines in the soft band that are compelling our fit to broaden the cooler component. It is clear that the asymmetric broad wings of these lines are the reason that a second broader \texttt{cie} is necessary. A shift between the hotter and cooler components is also evident. The effect of detector systematics on cooler gas velocity dispersion is discussed in \S\ref{sec:systematics} where we show that the systematics corrected velocity should still be $\sim$420 km s$^{-1}$.

Applying Eq. \ref{eq:bulk_velocity} for the cooler and hotter gas, we recover a velocity difference of $v_{\textrm{\scriptsize{bulk,hot-cool}}} = 450 \pm 140$ km s$^{-1}$ with the hotter gas redshifted and moving away from our line of sight. We also find some velocity difference between the cooler gas and the central galaxy, $v_{\textrm{\scriptsize{bulk, cool-CG}}} = -340 \pm 140$ km s$^{-1}$ (cooler gas blueshifted), but with only $2.4\sigma$ significance. Knowledge of detector systematics so far however indicates that this velocity difference may entirely be due to uncertainties in the energy-dependent instrumental gain correction (see \S\ref{sec:systematics}). Therefore, whether the cooler gas is experiencing a bulk motion with respect to the central galaxy as well needs to be established with more robust analysis of detector systematics and deeper data in the soft band for better statistics. The relative velocity difference between hotter and cooler gas can therefore be best interpreted as an upper limit. Bulk motion between hot gas and the central galaxy, on the other hand was found to be $v_{\textrm{\scriptsize{bulk, hot-CG}}} = 120 \pm 20$ km s$^{-1}$, i.e., similar to the single \texttt{cie} case. 

The measured S/Fe, Ar/Fe and Ca/Fe ratios are $1.2 \pm 0.2$ $Z_{\odot}$, $1.2 \pm 0.2$ $Z_{\odot}$ and $1.3 \pm 0.2$ $Z_{\odot}$, respectively. Thus, the data suggest that all these elemental ratios may be slightly super-solar, but still consistent with solar abundance ratios with the present errors. The Si/Fe ratio, on the other hand, was found to be $1.7^{+0.5}_{-0.3} Z_{\odot}$, which is super-solar at $2.3\sigma$ confidence level. Future deeper observations may be needed to conclusively establish or rule out super-solar abundance ratios in Cygnus A.

\begin{figure*}
    \centering
    \includegraphics[width=\textwidth, keepaspectratio]{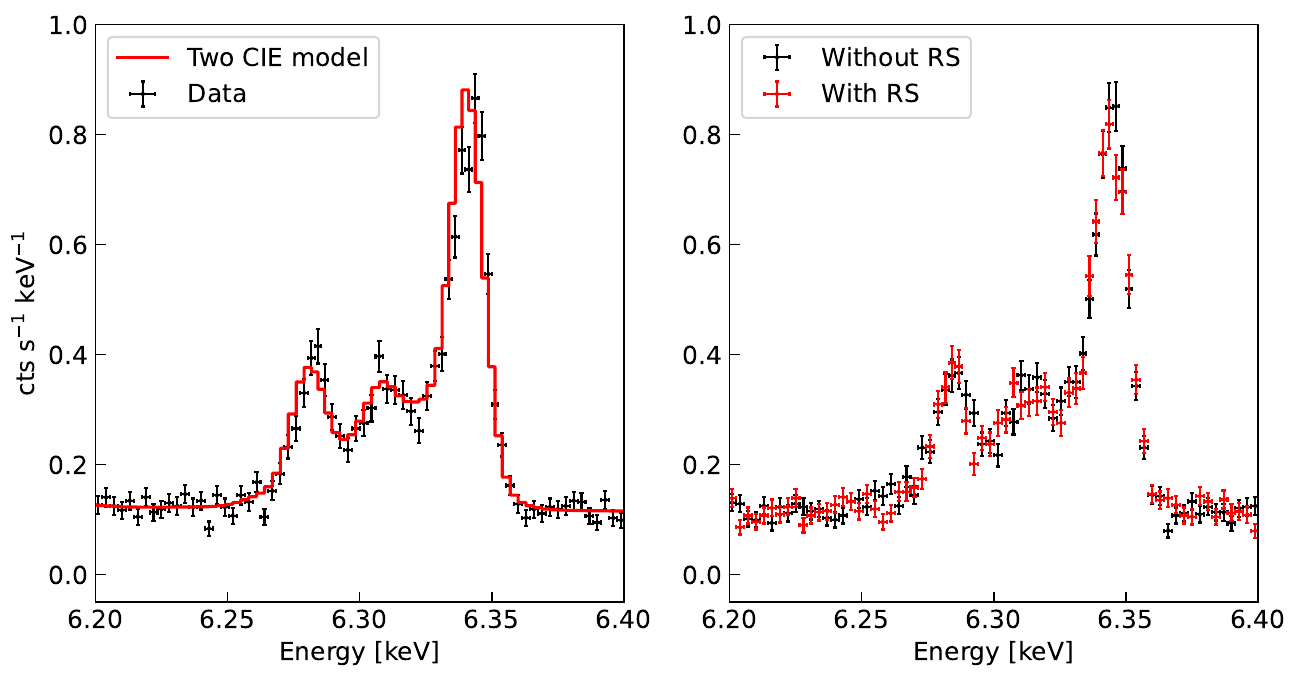}
    \caption{\small{\emph{Left:} Unfitted spectrum including the Fe-$w$ line in the $6.2-6.4$ keV band with the two \texttt{cie} model from \S\ref{subsec:two_cie} overlaid on top.} \emph{Right:} \texttt{clus} model simulation of the Fe-$w$ line for Cygnus A with and without resonance scattering (RS) for 170 ks.}
\label{fig:resonance_scattering}
\end{figure*}

\subsection{Comparison with \textit{Chandra}} \label{subsec:chandra}

Although \textit{Resolve} robustly detects the cooler component, we have also independently characterized this component with the \textit{Chandra} data. We followed this analysis as there are calibration uncertainties in the soft band of XRISM. Moreover, the \textit{Chandra} data covers the Fe-L complex which is sensitive to cooler gas. We chose ObsID 17512, taken with ACIS-I and has an observation start date of 15-09-2016, for our test as this is the deepest \textit{Chandra} pointing of the core of Cygnus A (68 ks). The data were reduced with standard procedures using \texttt{CIAO 4.17} \citep{2006SPIE.6270E..1VF} and \texttt{CALDB 4.17.2}. We extracted the spectrum in the XRISM field of view shown in Figure \ref{fig:fov} excluding the central 10 arcsec to avoid contamination from the AGN. We then fitted a two \texttt{cie} model absorbed with a neutral hydrogen column of $n_H = 4.14 \times 10^{21}$ cm$^{-2}$ and redshifted this model to $z = 0.0561$. The redshift and the $n_H$ were fixed during the fit. The normalization and temperature of each \texttt{cie} component were allowed to vary independently. The Fe abundance of the two components was coupled and then allowed to vary. All other elements were kept fixed at solar abundance. 

The spectrum in the $0.7-7.0$ keV band was then fitted after correcting for NXB using the ``stowed" background (see \citealt{majumder_decoding_2024} for more details). We find a $1.4 \pm 0.2$ keV component along with a $5.9 \pm 0.3$ keV component in this spectrum. Other fit parameters are noted in \S\ref{sec:chandra_fits}.  The effect of the decrease of soft-band effective area due to carbon-oxygen contamination in ACIS-I, on the other hand, is mostly limited to energies below 1 keV\footnote{\href{https://tinyurl.com/mur3kmwf}{ACIS QE Contamination}}. Therefore, the effect of this on our result should be small. This result confirms the cooler component seen in our XRISM data. If the cooler component temperature is fixed to 1.4 keV and the \textit{Resolve} spectrum is refit, the \texttt{cstat} increases by 4, but the fit parameters remain unchanged within their errors. The $\Delta$AIC, on the other hand, increases by 2, suggesting no strong statistical preference between a free cooler temperature versus fixing it to 1.4 keV. Therefore, the temperature measured by XRISM ($2.0 \pm 0.2$ keV) might be higher either due to calibration systematics or missing data below 1.7 keV. It is also noteworthy that, even if we fix the cooler temperature to the one determined by Chandra, i.e., 1.4 keV, the \textit{Resolve} abundance ratios remain modestly super-solar, hence this conclusion is robust to uncertainties in the thermal structure of the ICM.

\subsection{Resonance Scattering} \label{subsec:resonance_scattering}

We investigated Fe-$w$ line suppression due to resonance scattering by adding back the ignored line and overplotting the two \texttt{cie} model fits. The unfitted spectrum in the $6.2-6.4$ keV band is shown in Figure \ref{fig:resonance_scattering}. As can be seen from the unfitted Fe-$w$ line, the wings and the peak of the line match the model profile well. This indicates negligible resonance scattering in the core of Cygnus A. This result is surprising given that Cygnus A has high density and iron abundance at the core \citep{snios_cocoon_2018,majumder_decoding_2024} that is comparable to the Perseus cluster, where Fe-$w$ line suppression was seen \citep{hitomi_collaboration_measurements_2018}. However, the velocity broadening of Cygnus A is higher than Perseus, which can reduce the optical depth of scattering, leading to smaller line suppression. 

We investigated whether this result is expected or not by simulating a spherically symmetric model of Cygnus A with SPEX's \texttt{clus} model \citep{stofanova_clus_2025}.\footnote{\href{https://spex-xray.github.io/spex-help/models/clus.html}{https://spex-xray.github.io/spex-help/models/clus.html}} We used as input the density, 3D temperature, and 3D abundance profiles reported in \cite{majumder_decoding_2024}. The 3D profiles in that work were constructed from projected profiles similar to the profiles reported previously by \cite{snios_cocoon_2018}. The \texttt{clus} model parameters used for the simulation are reported in \S\ref{sec:clus_params}. We further used a constant turbulent velocity of 261 km s$^{-1}$ throughout the cluster volume for simplicity. In order to obtain the correct normalization of the total spectrum and compare with data, we added to the \texttt{clus} model a powerlaw contribution from the AGN. The entire model was then redshifted and absorbed through the galactic neutral hydrogen column. The AGN powerlaw, redshift and galactic neutral hydrogen columns were the same as those discussed and obtained in \S\ref{subsec:spec_model} and \S\ref{sec:results}. With this setup, we simulated a mock spectrum using observation RMF and ARF files for 170 ks exposure. We then extracted the spectrum within a radius of the \textit{Resolve} FOV and show it in the right panel of Figure \ref{fig:resonance_scattering} without and with resonance scattering.

The figure shows that the model predicts no noticeable suppression of the Fe-$w$ line in Cygnus A due to the high velocity dispersion, consistent with data. Therefore, the lack of resonance scattering in Cygnus A is in line with theoretical expectations. 

\section{Discussion} \label{sec:discussion}

\subsection{Non-thermal Pressure}

The velocity dispersion of the hotter gas in Cygnus A is low compared to the sound speed. The sound speed ($c_s$) can be calculated as

\begin{equation}
    c_s = \sqrt{\frac{\gamma kT}{\mu m_p}},
\end{equation}
\noindent where $\gamma = 5/3$ is the adiabatic index and $\mu = 0.61$ is the mean molecular weight. For a mean temperature  $kT = 5.54^{+0.17}_{-0.19}$ keV, we derive $c_s \sim 1200$ km s$^{-1}$. If the velocity dispersion is entirely due to isotropic turbulence, then the 3D Mach number can be determined as

\begin{equation}
    M_{3D} = \frac{\sqrt{\sigma_{3D}^2 + v_{\textrm{\scriptsize{bulk, hot-CG}}}^2}}{c_s} = \frac{\sqrt{3\sigma_{1D}^2 + v_{\textrm{\scriptsize{bulk, hot-CG}}}^2}}{c_s},
\end{equation}

\noindent where $\sigma_{3D}$ is the 3D velocity dispersion. With this equation, we derive $M_{3D} = 0.39 \pm 0.02$. However, the motion due to the cocoon shock has an unknown velocity component along the line of sight as well. Thus, part of the velocity dispersion is likely due to shock expansion rather than turbulence. Therefore, if part of the motion is anisotropic due to the shock, then the Mach number of the turbulence can be less than 0.39. 

An upper limit on the non-thermal pressure fraction for the gas can then be calculated as \citep{eckert_non-thermal_2019}

\begin{equation}
    \frac{P_{\textrm{\scriptsize{NT}}}}{P_{\textrm{\scriptsize{tot}}}} = \frac{M_{3D}^2}{M_{3D}^2 + 3/\gamma}.
\end{equation}

\noindent For the reported Mach number, the non-thermal pressure fraction is then $< 7.7 \pm 0.7\%$. This is consistent with less than $10\%$ of non-thermal pressure fraction seen in previous objects based on direct methods \citep{xrism_collaboration_xrism_2025-1,xrism_collaboration_xrism_2025,fujita_xrism_2025}, indirect methods like density power spectra \citep{zhuravleva_turbulent_2014,zhuravleva_gas_2018}, and based on universal ICM gas fraction (e.g., \citealt{eckert_non-thermal_2019}). 

For the cooler gas, on the other hand, the reported one-dimensional velocity dispersion of $440 \pm 130$ km s$^{-1}$ is significant compared to the local sound speed of $720^{+90}_{-50}$ km s$^{-1}$. This corresponds to a $M_{3D} = 1.1 \pm 0.3$, suggesting that the 3D Mach number may be greater than the sound speed for the cooler gas. Although again, the quoted $M_{3D}$ is an upper limit assuming a purely turbulent scenario.

\subsection{Energy Injection by the Radio Source}

These observations were intended to use the observed kinematics of the X-ray emission lines to assess the impact of the AGN on the ICM. Before XRISM \textit{Resolve}, the radio AGN 3C 84 in the Perseus Cluster was the only case where such measurements had been made \citep{hitomi_collaboration_quiescent_2016,hitomi_collaboration_measurements_2018}. These data showed that the ICM was surprisingly quiescent ($\sigma_{1D} <  200$ km s$^{-1}$). Similarly, low velocity dispersions have been measured by \textit{Resolve} in the Centaurus Cluster \citep{2025Natur.638..365X}, A2029 \citep{xrism_collaboration_xrism_2025-1}, and Hydra A \citep{rose_xrism_2025}.  In contrast, we have measured significantly broader lines in the case of Cygnus A, $\sigma_{1D} = 261$ km s$^{-1}$ for the hotter component and 440 km s$^{-1}$ for the cooler component. Given that Cygnus A's radio jets are much more powerful than those previously studied, this suggests that the unusual ICM kinematics are the result of energy injection by the radio source. Such large velocities indicate that we may be detecting the expanding phase of the central cavities.

To quantify this, we have combined the kinematic data provided by \textit{Resolve} (Table \ref{tab:ICM_fits}) with the detailed analysis of the structure and physical conditions of the ICM around Cygnus A made by \cite{snios_cocoon_2018}.  To do so, we assume that the broader component in the emission lines produced by the cooler gas arises in the cooler central region of the cluster \citep{snios_cocoon_2018} and that the narrower component arises in the hotter gas that surrounds the central region. The central region is described by $n_e = 0.04$ cm$^{-3}$ within 35 kpc and the outer surrounding region by $n_e \propto  r^{-1.6}$ between 35 and 100 kpc, i.e., out to the edge of the \textit{Resolve} field of view.  The resulting values of $n_en_pV$ are 7 $\times 10^{66}$ cm$^{-3}$ for the central region and 2 $\times 10^{67}$ cm$^{-3}$ for the surrounding region with the central region consistent with the two-temperature fit of the XRISM data (Table \ref{tab:ICM_fits}). The fitted normalization in the surrounding region from XRISM is slightly higher than the above estimate, but this is expected since the fitted normalization accounts for extra emission from the corners of the detector and any scattering in and out of the field of view. Nevertheless, the consistency between normalizations based on \textit{Chandra} imaging and XRISM spectra validates our identification of the broader and narrower emission line components in Table \ref{tab:ICM_fits} with the central and the surrounding outer region in the \cite{snios_cocoon_2018} data, respectively. 

The gas masses are then 2.5 $\times 10^{11} M_{\odot}$ for the central region and 1.8 $\times 10^{12}$ $M_{\odot}$ for the surrounding outer region. The kinetic energy can then be derived as 

\begin{equation}
    E_{\textrm{\scriptsize{kinetic}}} = \frac{3}{2}M\sigma^2_{1D},
\end{equation}

\noindent where $M$ is the gas mass in a respective region. We derive $E_{\textrm{\scriptsize{kinetic}}} = 1.4 \times 10^{60}$ erg for the central region and $3.7 \times 10^{60}$ erg in the surrounding outer region. The total kinetic energy is therefore $5.1 \times 10^{60}$ erg. This value is very close to the \cite{snios_cocoon_2018} estimate of $4.7 \times 10^{60}$ erg for the total energy associated with the Cygnus A radio source, suggesting that the high velocities are due to the interaction of the radio source with the ICM. 

The morphology of the X-ray emission in this inner region is complex (see Figure \ref{fig:fov} left panel), with numerous filaments (called the X-ray ribs by \citealt{duffy_x-ray_2018}). This contrasts with the much smoother X-ray morphology at larger radii. Such a dichotomy in structure further supports the idea that the inner region is physically and dynamically distinct from the rest of the ICM. \cite{duffy_x-ray_2018} suggests that this inner region has resulted from the disruption of the ICM's cool-core by the radio source, which would be consistent with our kinematic results.

\subsection{Timescales and Kinetic Powers} \label{subsec:timescales}

In addition to assessing the total kinetic energy in the gas, it is also important to understand the timescale over which this energy was delivered and compare the resulting kinetic power to the rate at which the gas in this region is cooling through X-ray emission.

Given the agreement between the energy associated with the radio source and the ICM kinetic energy, the most natural timescale would be the lifetime of the radio source. \citet{snios_cocoon_2018} estimate an age of 20 Myr, which would then imply that kinetic energy has been transferred to the gas at a rate of $7.4 \times 10^{45}$ erg s$^{-1}$.

Another possible timescale would be a simple crossing-time estimate taken to be the radius of the relevant region of the ICM divided by $\sigma_{3D}$. For the inner region, this would be $t_{\textrm{\scriptsize{cross}}}$ = 35 kpc/760 km s$^{-1}$ = 45 Myr, while for the outer region it would be $t_{\textrm{\scriptsize{cross}}}$ = 100 kpc/450 km s$^{-1}$ = 216 Myr. At least for the outer region, the timescale is much longer than the radio source age, and is only possible if the source of kinetic energy in this gas is related to an older outburst in Cygnus A. Possible signatures of past outbursts, including one older than $200$ Myr, were reported in \cite{majumder_decoding_2024}. If we assume a timescale of 216 Myr, we can place a lower limit of $6.9 \times 10^{44}$ erg s$^{-1}$ for the rate at which kinetic energy has been delivered to the gas.

The cooling luminosity of the X-ray gas within 100 kpc, on the other hand, is $4 \times 10^{44}$ erg s$^{-1}$ \citep{rafferty_feedback-regulated_2006}, far below the kinetic power based on the radio source lifetime and less than that based on the crossing time. Provided that much of the kinetic energy is converted to thermal energy, this implies that the ICM around Cygnus A is far from heating and cooling equilibrium. 


\section{Summary} \label{sec:summary}

We analyzed a 170 ks net exposure of the core of Cygnus A using XRISM \textit{Resolve}. Our findings can be summarized as follows:

\begin{itemize}
    \item For a one thermal collisional ionization equilibrium (\texttt{cie}) model, spectral fit results suggest that the one-dimensional (1D) velocity dispersion of the $5.33 \pm 0.11$ keV cluster gas is $273 \pm 13$ km s$^{-1}$. Furthermore, the gas is also experiencing a bulk velocity motion of $100 \pm 20$ km s$^{-1}$ (ICM redshifted) with respect to the central galaxy along the line of sight.
    \item The spectrum, however, is a better fit with two \texttt{cie} models. In this case, we report that the hotter gas of $5.53 \pm 0.13$ keV, has a velocity dispersion of $261 \pm 13$ km s$^{-1}$ and a relative velocity of $120 \pm 20$ km s$^{-1}$ with respect to the central galaxy (hotter gas redshifted) along the line of sight, i.e., similar to the single \texttt{cie} case. The cooler gas component of $2.0^{+0.4}_{-0.3}$ keV, however, has an even broader velocity dispersion of $440 \pm 130$ km s$^{-1}$ ($3.4\sigma$ significance). The velocity broadening in Cygnus A is thus one of the largest published so far among the clusters observed by XRISM. As a result, little suppression of the $w$-line in the Fe XXV He$\alpha$ line complex due to resonance scattering is found. 
    \item If the velocity broadening is purely turbulence (unlikely), then the turbulent three-dimensional (3D) Mach number is $0.39 \pm 0.02$ for the hotter gas and $1.1 \pm 0.3$ for the cooler gas. The cooler gas Mach number is thus transonic. However, in case the cocoon shocks of Cygnus A are contributing to the velocity broadening (likely), the turbulent Mach numbers are less than these reported values. 
    \item The measured Si/Fe, S/Fe, Ar/Fe and Ca/Fe abundance ratios are $1.7^{+0.5}_{-0.3} Z_{\odot}$, $1.2 \pm 0.2$ $Z_{\odot}$, $1.2 \pm 0.2$ $Z_{\odot}$ and $1.3 \pm 0.2$ $Z_{\odot}$, respectively. Thus, the data suggest that S to Ca ratios may be slightly super-solar, although still consistent with solar ratios within errors. Present measurements do however suggest that the Si/Fe ratio may be super-solar ($2.3\sigma$ significance). The modestly super-solar ratios measured by \textit{Resolve} remain robust even if the temperature of the cooler component is as low as 1.4 keV, as suggested by the \textit{Chandra} data which includes the Fe-L emission from cooler ICM.
    \item The upper limit on the non-thermal pressure contribution by the hotter gas in Cygnus A is $7.7 \pm 0.7\%$. This is consistent with $< 10\%$ non-thermal pressure fraction seen in previous studies based on direct \citep{xrism_collaboration_xrism_2025-1,xrism_collaboration_xrism_2025,fujita_xrism_2025} or indirect methods \citep{zhuravleva_turbulent_2014,zhuravleva_gas_2018}, and based on universal ICM gas fraction (e.g., \citealt{eckert_non-thermal_2019}). The non-thermal pressure fraction is nevertheless higher than what is reported so far for clusters with XRISM. 
    \item We associate the cooler broad \texttt{cie} component with the cooler central region ($0-35 $ kpc) of the cluster \citep{snios_cocoon_2018} and the hotter narrow \texttt{cie} component with the gas surrounding the inner central region ($35-100$ kpc). The expected spectral normalizations from such theoretical assumptions agree well with what is observed from the spectral fits. The kinetic energy associated with the cooler gas is then $1.4 \times 10^{60}$ erg and $3.7 \times 10^{60}$ erg for the hotter gas. The total kinetic energy of $5.1 \times 10^{60}$ erg is very close to the \cite{snios_cocoon_2018} estimate of $4.7 \times 10^{60}$ erg for the total energy associated with the Cygnus A radio source. Thus, the observed velocities are almost certainly due to the interaction of the radio source with the cluster gas.
    \item The power at which energy is being injected into the cluster gas can be estimated by either assuming a radio age of $20$ Myr \citep{snios_cocoon_2018} or the crossing timescale of $45$ Myr for cooler gas, and $216$ Myr for the hotter gas (see \S\ref{subsec:timescales}). The kinetic energy injection is then $6.9 \times 10^{44} - 7.4 \times 10^{45}$ erg s$^{-1}$, assuming the actual timescale is between radio age and crossing timescales. This power range is higher than the cooling luminosity of $4 \times 10^{44}$ erg s$^{-1}$ within $\sim 100$ kpc of Cygnus A \citep{rafferty_feedback-regulated_2006}. Assuming that the kinetic energy is efficiently converted to thermal energy, our results suggest that the gas around Cygnus A departs from heating and cooling equilibrium.
\end{itemize}
\vspace{-0.7cm}
\begin{acknowledgments}

AM thanks Dr. Daniela Huppenkothen (UvA) for discussions on statistical modeling of the data and Dr. Jelle de Plaa (SRON) for help with SPEX model fitting. TH acknowledges the financial support from NASA 80NSSC25K7537 for this project. AM, AS, LG and MW acknowledge support from the Netherlands Organisation for Scientific Research (NWO). BRM acknowledges support from the Canadian Space Agency and the National Science and Engineering Research Council of Canada. This research has made use of data obtained from the XRISM data archive maintained by NASA HEASARC and JAXA DARTS. The research also made use of the \textit{Chandra} data archive provided by the \textit{Chandra} X-ray Center (CXC). 

\end{acknowledgments}

\section*{Data availability}

The XRISM raw data used in this work will become publicly available following a one-year embargo period that ends at the end of October 2025. This paper employs a list of Chandra datasets, obtained by the Chandra X-ray Observatory, contained in ~\dataset[https://doi.org/10.25574/cdc.492].. All intermediate data products and scripts used to obtain the results are publicly available at \cite{maj25-zen}.

\facilities{XRISM (\textit{Resolve}), \textit{Chandra} (ACIS)}

\software{\texttt{NUMPY} \citep{vdw11}, 
          \texttt{ASTROPY} \citep{astropy:2013,astropy:2018,astropy:2022},
          \texttt{MATPLOTLIB} \citep{hun07},
          \texttt{APLPY} \citep{rob12}}

\appendix
\vspace{-0.5cm}

\section{Fits with Ignored Lines} \label{appendix:ig_lines}

\begin{figure}[h!]
    \centering
    \includegraphics[width=0.4\columnwidth, keepaspectratio]{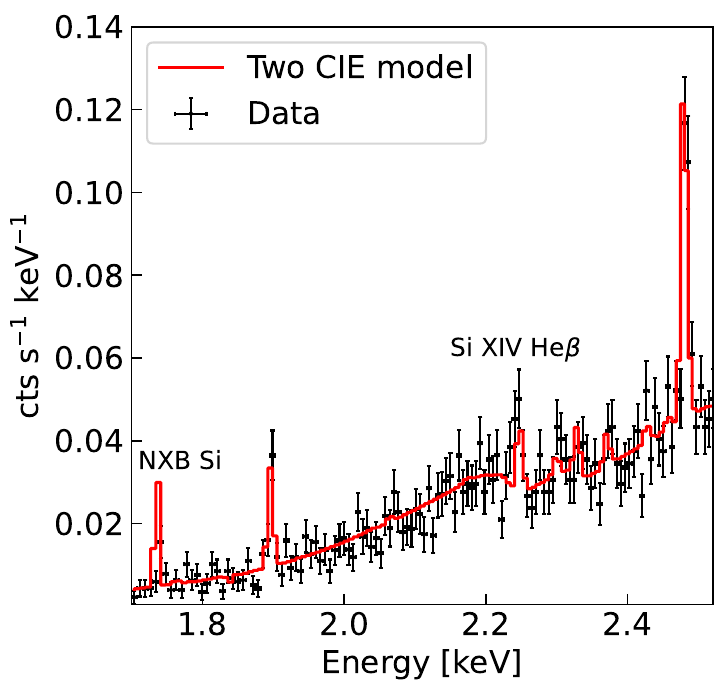}
    \caption{\small{Two \texttt{cie} model fit to the data in the $1.7-12.0$ keV band. We show the $1.7-2.5$ keV band where the Si NXB Fluorescence line and Si He-$\beta$ has been used in the fitting. The data has been binned by a factor of 2 for visual purposes.}}
\label{fig:calibration}
\end{figure}

In \S\ref{sec:results}, we ignored the Si Fluorescence instrumental line in the $1.735-1.75$ keV band as well as the Si He-$\beta$ line in the $2.22-2.265$ keV band. We show the fit with these two lines included in Figure \ref{fig:calibration}. As can be seen, both lines are poorly modeled. The model line overshoots the Fluorescence line, most likely due to calibration issues with this particular NXB line. The Si He-$\beta$, on the other hand, falls at the Si edge of the effective area at Cygnus A's redshift, and the fit is thus unable to fit this line properly. The \texttt{cstat} of this fit is $3113$ and is substantially worse than the fits reported in \S\ref{sec:results}. To avoid biasing the spectral fits, we therefore ignored these lines.  

\FloatBarrier

\section{Effect of detector systematics on the cooler CIE component} \label{sec:systematics}

Due to XRISM's gate valve closed configuration, the Resolve instrument team is yet to undertake an in-flight pixel-by-pixel energy-scale calibration of the detector in the soft band ($< 5.4$ keV; for the $5.4-9.0$ keV band, see the discussion in \S\ref{sec:obs_data}). The current recommendation from the instrument team is therefore to assume a conservative systematic uncertainty of $\pm 1$ eV on the energy-scale accuracy in the soft band. The $+1$ eV offset of the Si instrumental line in stacked source data necessitates such a recommendation. At 2.5 keV, this uncertainty corresponds to $\pm 120$ km s$^{-1}$, less than the bulk velocities for the cooler gas reported in \S\ref{subsec:two_cie}. 

Furthermore, the line-spread function (LSF) can also be affected due to this uncertainty. Energy-scale misalignment across pixels or across times in the integrated data can propagate energy-scale errors to LSF errors. Energy scales between 6-7 keV should be well aligned since the scales are pinned at 5.9 keV. However, at lower energies, a dispersion correction due to propagated systematic errors on LSF is needed. If the conservative energy-scale errors induce a corresponding 1 eV broadening in the soft-band, then a measured velocity dispersion of 120 km s$^{-1}$ at 2.5 keV would be consistent with 0. The velocity dispersion of the cooler gas reported in \S\ref{subsec:two_cie} is significantly higher than this. Simple quadrature subtraction of 120 km s$^{-1}$ from the velocity dispersion results in $\sim$420 km s$^{-1}$ which is still significant. We do however note that the simple quadrature subtraction used to estimate corrected velocity dispersion is not accurate if the energy dependent shift varies in magnitude with time. At this time, it is unknown if this is the case. Future robust studies of soft band systematics from the Resolve instrument team is required to better understand the systematics. We also check the effect on bulk velocity by manipulating the redistribution matrix file (RMF). We shift the RMF by:
\begin{equation}
    \begin{split}
        (2.5 - \textrm{shift}) = 2.5 A + B \\
        6.7 = 6.7A + B,
    \end{split}
\end{equation}

\noindent where, shift = $\pm$0.001 keV. Using these equations, we obtain $A = 1.00024$, $B = -0.001595$ for shift $= 0.001$ keV and $A = 0.99976$, $B = 0.001595$ for shift $=-0.001$ keV. Using these modified RMFs, we refitted the full field of view Resolve spectrum using the two CIE + AGN models. For a shift of $+0.001$ keV, we obtained cooler gas velocity of $-480 \pm 140$ km s$^{-1}$ with respect to the central galaxy while for a shift of $-0.001$ keV, we obtained a cooler gas velocity of $-50 \pm 140$ km s$^{-1}$ with respect to the central galaxy.

Thus, although the velocity dispersion result from our analysis should be robust, it is possible that the cooler gas bulk motion with respect to the central galaxy we derive can be entirely due to systematics, if the energy-scale uncertainty in the soft band is indeed as high as 1 eV.

\section{\textit{Chandra} Fits} \label{sec:chandra_fits}

We note in Table \ref{tab:chandra_fits} the fit parameters for the \textit{Chandra} spectrum extracted from the XRISM field of view (see Figure \ref{fig:fov}). The central 10 arcsec around the central AGN was excluded to avoid contamination to the cluster emission. Only the parameters that were varied are noted in the table. The fixed parameters are described in \S\ref{subsec:chandra}.

\begin{table}[h!]
    \centering
\caption{Best-fit parameters for Cygnus A in the $0.7-7.0$ keV band of \textit{Chandra}.} 
\label{tab:chandra_fits}
    \begin{tabular}{cc}
    \hline
    \hline
    Parameter&Value\\
    \tableline
    \multicolumn{2}{c}{Model $-$ \texttt{reds(hot(cie+cie))\tablenotemark{a}}}\\
        Hotter gas norm ($n_e n_p V$)&$1.75^{+0.05}_{-0.06} \times 10^{67}$ cm$^{-3}$\\
        Hotter gas kT&$5.9 \pm 0.3$ keV\\
        Cooler gas norm ($n_e n_p V$)&$0.16^{+0.07}_{-0.05} \times 10^{67}$ cm$^{-3}$\\
        Cooler gas kT&$1.4 \pm 0.2$ keV\\
        Fe &$0.80 \pm 0.03$ $Z_{\odot}$\\
 \tableline
    \end{tabular}
    \tablecomments{$n_e$ is the electron density, $n_p$ is the proton density and $V$ is the line-of-sight enclosed volume. Only the parameters that were varied are noted in the table. The fixed parameters are described in \S\ref{subsec:chandra}.}
    \tablenotetext{a}{A detailed spectral analysis should also include any non-thermal emission from the lobes and jets of Cygnus A (e.g., see \citealt{devries_detection_2018}). Such an analysis is beyond the scope of this work.}
\end{table}

\FloatBarrier

\section{\texttt{clus} Model Parameters} \label{sec:clus_params}

We report the parameters of the \texttt{clus} model\footnote{\href{https://spex-xray.github.io/spex-help/models/clus.html}{https://spex-xray.github.io/spex-help/models/clus.html}} in Table \ref{tab:clus_params} with which the simulations were done in \S\ref{subsec:resonance_scattering}. The density, temperature and abundance profile with which these parameters were obtained are reported in \cite{majumder_decoding_2024}. 

\begin{table}[h!]
    \centering
\caption{\texttt{clus} model parameters for Cygnus A simulation.} 
\label{tab:clus_params}
    \begin{tabular}{cccc}
    \hline
    \hline
    Parameter & Value & Parameter & Value\\
    \tableline
        $n_{0,1}$ & 90070 m$^{-3}$ & A & 4.17\\
        $r_{tc}$ & $0.017r_{500}$ & B & $0.048r_{500}$\\
        $\beta_1$ & 0.595 & C & 1.73\\
        $n_{0,2}$ & 0 m$^{-3}$ & D & $0.74$\\
        $T_c$ & 2.85 keV & E & $1.4 \times 10^{-3}r_{500}$\\
        $T_h$ & 10 keV & F & $100r_{500}$\\
        $r_{\textrm{to}}$ & $0.056r_{500}$ & G & 0.23\\
        $\mu$ & 1.16 & $r_{500}$ & $3.8$ ($10^{22}$ cm)\tablenotemark{a}\\
        r$_{\textrm{c,2}}$ & $0.44r_{500}$ & $r_{out}$ & 1\\
        a & $10^{-11}$ & $r_{min}$ & 1\\
        b & $10$ & $r_{max}$ & 0.079\tablenotemark{b}\\
        c & 3.24 & $a_v$ & 261$\sqrt{2}$ km s$^{-1}$ \tablenotemark{c}\\
 \tableline
    \end{tabular}
    \tablenotetext{a}{$r_{500} = 1235$ kpc (1134 arcsec; \citealt{halbesma_simulations_2019})}
    \tablenotetext{b}{For \textit{Resolve} field of view radius of $\sim 90$ arcsec divided by $r_{500} = 1134$ arcsec and $r_{out}=1$.}
    \tablenotetext{c}{In \texttt{clus} model, the velocity parameter describes the magnitude of micro-turbulence, which is equal to $\sqrt{2}$ times 1D velocity dispersion.}
\end{table}

\FloatBarrier

\pagebreak

\bibliographystyle{aasjournalv7}

\end{document}